\documentclass[12pt]{article}

\usepackage{amsmath,amssymb,amsthm}

\newcommand{\vtri}{\boldsymbol{\vartriangle}}

\begin{document}

\title{\bf  
Interacting gauge fields and 
the zero-energy eigenstates in two dimensions %in Gel'fand triplets
}

\author{Tsunehiro Kobayashi\footnote{E-mail: 
kobayash@a.tsukuba-tech.ac.jp} \\
{\footnotesize\it Research Center on the Higher Education 
for the Hearing and visually Impaired,}
{\footnotesize\it Tsukuba College of Technology}\\
{\footnotesize\it Ibaraki 305-0005, Japan}}

\date{}

\maketitle

\begin{abstract}

Gauge fields are formulated 
in terms of the zero-energy eigenstates of 2-dimensional 
Schr$\ddot {\rm o}$dinger equations with central potentials 
$V_a(\rho)=-a^2g_a\rho^{2(a-1)}$ ($a\not=0$, $g_a>0$ and  
$\rho=\sqrt{x^2+y^2}$). 
It is shown that 
the zero-energy states can naturally be interpreted as a kind of 
interacting gauge fields of which effects are solved as the factors 
$e^{ig_c\chi_A}$, where $\chi_A$ are complex gauge functions written by 
the zero-energy eigenfunctions. 
We see that the gauge 
fields for $a=1$ are nothing but 
tachyons that have negative squared-mass $m^2=-g_1$. 
We also find out U(1)-type gauge fields for $a=1/2$  and 
SU(3)-type gauge fields for $a=3/2$. 
Massive particles with internal structures 
described by the zero-energy states 
 are also studied.  

\end{abstract}

\thispagestyle{empty}

\setcounter{page}{0}

\pagebreak

It has been shown that 
2-dimensional Schr$\ddot {\rm o}$dinger equations with the central potentials 
$V_a(\rho)=-a^2g_a\rho^{2(a-1)}$ ($a\not=0$ and  
$\rho=\sqrt{x^2+y^2}$, i.e., $x=\rho \cos \varphi$ and $y=\rho \sin \varphi$.)
have zero-energy eigenstates which are infinitely degenerate~
\cite{sk-jp,k-1,ks-pr,k-ps}. 
Let us briefly repeat the argument for driving the zero-energy states.
The Schr$\ddot {\rm o}$dinger equations 
for the energy eigenvalue ${\cal E}$, which 
are written as
\begin{equation}
 [-{\hbar^2 \over 2m}\vtri(x,y) +V_a(\rho)]\ \psi(x,y) 
  = {\cal E}\ \psi(x,y)
  \label{0}
\end{equation}
where 
$%$
\vtri(x,y)=\partial^2/ \partial x^2+\partial^2 / \partial y^2,
$  
have zero-energy (${\cal E}=0$) eigenstates. 
Note here that in this equation the mass $m$ and the coordinate vector $(x,y)$ 
can represent not only those of the single particle but those of the center of 
mass for a many particle system as well~\cite{k-ps}. 
It has also been shown that, 
as far as the zero-energy eigenstates ($\psi_0$) are concerned, 
the Schr$\ddot {\rm o}$dinger equations for all $a$ can be reduced to 
the following equation 
in terms of the conformal transformations 
$\zeta_a=z^a$ with $z=x+iy$~\cite{k-1,ks-pr}; 
\begin{equation}
[-{\hbar^2 \over 2m}{\boldsymbol{\triangle}}_a-g_a]\ \ \psi_0 (u_a,v_a)=0, 
 \label{1}
\end{equation}
where
$ 
{\boldsymbol{\triangle}}_a=\partial^2/ \partial u_a^2+\partial^2/ \partial v_a^2,
$  
using the variables defined by 
the relation $\zeta_a=u_a+iv_a$, where $u_a=\rho^a \cos a\varphi$ and 
$v_a=\rho^a \sin a\varphi$. 
That is to say, the zero-energy eigenstates 
for all the different numbers of $a$ 
are described by the same plane-wave 
solutions in the $\zeta_a$ plane. 
Furthermore it is easily shown 
that the zero-energy states degenerate infinitely.
Let us consider the case for $a>0$ and $g_a>0$. 
Putting the function $f^\pm_n (u_a;v_a)e^{\pm ik_a u_a}$ with 
$k_a=\sqrt{2mg_a}/\hbar $ 
into Eq.~\eqref{1}, 
where $f^\pm_n (u_a;v_a)$ are polynomials of degree $n$ ($n=0,1,2,\cdots $), 
%functions of $u_a$ and $v_a$, 
we obtain the equations for the polynomials 
\begin{equation}
[{\boldsymbol{\triangle}}_a \pm2ik_a{\partial \over \partial u_a}]
f^\pm_n (u_a;v_a)=0.
\label{2}
\end{equation} 
Note that from the above equations we can easily see the relation  
$(f^-_n (u_a;v_a))^*=f^+_n (u_a;v_a)$ for all $a$ and $n$. 
General forms of the polynomials have been obtained by using the solutions 
in the $a=2$ case 
(2-dimensional parabolic potential barrier)~\cite{k-1,ks-pr}. 
Since all the solutions have the factors $e^{\pm ik_a u_a}$ 
or $e^{\pm ik_a v_a}$, we see that 
the zero-energy states describe stationary flows~\cite{sk-jp,k-1,ks-pr}. 
Taking account of the direction of incoming flows in the 
$\zeta_a$ plane that is expressed by the angle $\alpha$ 
to the $u_a$ axis, the general eigenfunctions 
with zero-energy are written as arbitrary linear combinations of 
eigenfunctions included in two infinite 
series of $\left\{ \psi_{0n}^{\pm (u)}(u_a(\alpha);v_a(\alpha))  \right\}$ 
%and $\left\{ \psi_{0n}^{\pm (v)}(v_a(\alpha);u_a(\alpha))  \right\}$ 
for $n=0,1,2,\cdots$, where 
%\end{document}
%$$ 
\begin{equation} 
\psi_{0n}^{\pm (u)}(u_a(\alpha);v_a(\alpha))=
f_n^\pm(u_a(\alpha);v_a(\alpha))e^{\pm ik_a u_a(\alpha)}, %\nonumber \\ 
\label{psi}
\end{equation}
$u_a(\alpha) =u_a {\rm cos}\alpha +v_a {\rm sin}\alpha $, 
and $0 \leq \alpha < 2\pi$. 
%$v_a(\alpha) =v_a {\rm cos}\alpha -u_a {\rm sin}\alpha$. 
(For the details, see the sections II and III of Ref.~\cite{ks-pr}.)
It has been also pointed out that  
the motions of the $z$ direction perpendicular to the $xy$ plane 
can be introduced as 
free motions represented by $e^{\pm ik_z z}$. 
In this case 
the total energies $E_T$ of the states are given by 
$E_T=E_z$, 
where $E_z=(\hbar k_z)^2/2m$ are 
the energies of the free motions in the $z$ direction. 
Note that the zero-energy eigenfunctions can not be normalized 
as same as those 
in scattering processes~\cite{bohm}. 
Actually it has been shown that all the zero-energy states 
for $g_a>0$ are eigenstates 
in the conjugate spaces of the Gel'fand triplets~\cite{sk-jp,k-1,ks-pr,sk-nc}. 
How we can understand the infinite degeneracy of the zero-energy states 
is the problem discussed in this letter.

Let us start our discussion from the equation in three dimensional spaces 
\begin{equation}
 [-\vtri(x,y,z) %+{\partial^2 \over \partial z^2}) 
 +V_a(\rho)]\ \psi({\bf r}) %e^{ip_z z/\hbar}
  = {\cal Q}\ \psi({\bf r}), %e^{ip_z z/\hbar},
  \label{F}
\end{equation}
where $\vtri(x,y,z)=\vtri(x,y)+\partial^2 /\partial z^2$. 
Note that the dimension of the potentials $ V_a$ are different from 
those in Eq.~\eqref{0}, and that of ${\cal Q}$ is also different from 
the energy dimension. 
We consider the zero-energy solutions for the $xy$ plane and 
the free motion with the non-zero momentum $p_z$ in the $z$ direction. 
This means that ${\cal Q}=(p_z/\hbar)^2$. 
We notice that the solutions have the specific direction $z$ and 
then the wave functions should be expressed 
by the $z$ component of a vector such that 
%${\bf \psi}({\bf r})=(0,0,\psi_z({\bf r}))$ where 
$$
\psi_z({\bf r})=A_z(x,y)e^{ip_z z/\hbar}.
$$
 Note that in general we can write Eq.~\eqref{F} in a rotationally 
 symmetric form in terms of the totally anti-symmetric symbols 
 $\epsilon_{ijk}$ defined by $\epsilon_{123}=1$ in three dimensions 
 as follows; 
 \begin{equation}
 [-(\epsilon_{ijk}\partial_j {\hat p}_k 
 \epsilon_{ilm}\partial_l {\hat p}_k+
 ({\hat p}\cdot \partial)^2)
  +V_a(\rho)]\ \psi_{\hat p}({\bf r}) 
  = {\cal Q}\ \psi_{\hat p}({\bf r}),
  \label{F1}
\end{equation}
where $\hat p={\bf p}/p$ with $p=|{\bf p}|$ stands for 
the unit vector of the direction of the non-zero momentum ${\bf p}$, 
and $\rho^2=\epsilon_{ijk}r_j{\hat p}_k \epsilon_{ilm}r_l {\hat p}_m$ 
should be taken. 
Note also that Eq.~\eqref{F1} depends only on the direction of the 
momentum but not on the magnitude of the momentum. 
Hereafter we choose the $z$ direction as the specific direction 
of the momentum. 

Let us introduce the vector fields ${\bf A}_n^{\pm0}e^{ip_z z/\hbar}$ 
which have the specific direction $z$ 
are the eigenstates of Eq.~\eqref{F1} for ${\cal Q}=(p_z/\hbar)^2$, 
where 
${\bf A}_n^{\pm0}=(0,0,A_z^\pm(x,y))$ and 
$
  A_z^\pm(x,y)_n=\psi_{0n}^\pm
$ for ($n=0,1,2,\cdot \cdot \cdot$) 
and $\psi_{0n}^\pm$ are obtained by replacing 
$k_a=\sqrt{2mg_a}/\hbar$ 
with $k_a=\sqrt{g_a}$ in the zero-energy eigenstates 
given by Eq.~\eqref{psi}. 
Therefore the relations $({\bf A}_n^{-0})^*={\bf A}_n^{+0}$ hold. 
We study the property of the vector fields ${\bf A}_n^\pm$ 
\begin{equation}
{\bf A}_n^\pm=(\partial_x \chi^\pm(xy)_n,\partial_y \chi^\pm(x,y)_n,
A_z^\pm(x,y)_n),
\label{A}
\end{equation}   
where 
the functions $\chi_n$ are defined by 
\begin{equation}
\chi^\pm(x,y)_n=z A_z^\pm(x,y)_n.
\label{chi}
\end{equation} 
 Note that $\chi_n^\pm$ is generally expressed in rotational invariant forms 
${\bf r}\cdot {\bf A}_n^{\pm0}$, and we have a relation 
$${\bf A}_n^\pm={\bf \partial}\chi_n^\pm.$$ 
In order to derive the last relation the fact that $A_z(x,y)$ do not depend 
on $z$ at all is important. 
We easily see that $\chi_{n}$ satisfy the same equation as $A_{zn}^\pm $ 
\begin{equation} 
[\vtri(x,y,z)-V_a(\rho)]\chi_{n}^\pm=0. 
\label{G2}
\end{equation} 
It is obvious that the fields also satisfy the equation 
\begin{equation}
 [-\vtri(x,y,z) 
 + V_a(\rho)]{\bf A}(x,y)_n^\pm
  = 0 .  
  \label{F0}
\end{equation}
Hereafter we take off the $\pm$ symbols from ${\bf A}_n^\pm$, $\chi_n^\pm$ 
and so on. 
Now we introduce scalar functions $\chi_{nm}=\chi_m-\chi_n$ 
for $n, \ m=0,1,2,3,\cdot\cdot\cdot$.  
We easily see that 
\begin{equation} 
 A_{zn}+\partial_z\chi_{nm}=A_{zm}, 
\label{G1}
\end{equation} 
and thus we have reltions 
\begin{equation} 
 {\bf A}_{n}+{\bf \partial}\chi_{nm}={\bf A}_{m}. 
\label{G0}
\end{equation} 
It is trivial that these conditions can be extended to 
 arbitrary suitable pairs of 
 linear combinations of $A_n$ and $\chi_{nm}$. 
Thus we see that the transformation~\eqref{G0} in terms of $\chi$ 
is a kind of the gauge transformation for the vector fields ${\bf A}_n$ 
interacting with the external potentials $V_a$. 
In this picture the freedom of the gauge transformations can be understood 
as the freedom arising from the infinite degeneracy of the zero-energy 
states in two dimensions. 
It is also quite naturally understood 
that the gauge fields are represented by the transverse waves 
perpendicular to the moving direction. 
Now we may say that the vector fields ${\bf A}$ 
are a kind of gauge fields that are 
interacting with the external potentials $V_a$. 
It should be strongly noticed that the fields ${\bf A}$ are not always real 
rather complex in general. 
Actually the polynomials $f_{n}^{\pm }$ given in Eq.~\eqref{psi} 
are real only for $n=0,1$ and complex for $n\geq 2$
~\cite{sk-jp,k-1,ks-pr,k-ps}.

Let us here proceed from Schr\"odinger equations in 3-dimensions 
to Klein-Gordon equations for massless particles 
in 4-dimensional Lorentz spaces. 
To make the discussion simple, we consider the case 
where a particle is moving in the $z$ direction with the non-zero momentum 
$p_z$. 
We may write it as 
\begin{equation} 
[{1 \over c^2}{\partial^2 \over \partial t^2} 
- \vtri(x,y,z)+V_a(\rho)]\psi_z(r)=0,
 \label{rel} %\label{E}
\end{equation} 
where $\psi_z(r)=A_z(x,y)e^{i(p_zz-p_ot)/\hbar}$. 
Since $p_0^2=p_z^2$ for the massless particle, 
the derivatives with respect to the $z$ and $t$ 
cancel out each other in Eq.~\eqref{rel}, and thus we have the same equation 
for $A_z$ as that for the zero-energy eigenstates.  %of Eq.~\eqref{F0}. 
Now we can introduce the 4-vector fields as 
$$ A_n= (A_0,{\bf A}_n),$$ 
where $A_0$ should be taken as functions satisfied by 
$\partial_t A_0=0$, and actually we may put $A_0=0$. 
In general the separation of 
the two directions perpendicular to 
the non-zero 3-momentum ${\bf p}$ in 4-dimensional spaces can be carried out 
in terms of the 4-momenta defined by $p^+=(|{\bf p}|,{\bf p})$ 
and $p^-=(-|{\bf p}|,{\bf p})$ such that 
$
\epsilon^{\mu \nu \lambda \sigma} p^+_\lambda p^-_\sigma,
$
where $\epsilon^{\mu \nu \lambda \sigma}$ is 
the totally anti-symmetric tensor 
defined by $\epsilon_{0123}=1$. 
It is trivial to write $\rho=x^2+y^2$ and 
$
\partial^2 / \partial x^2 + \partial^2 / \partial y^2 
$ 
in the covariant expressions in terms of these tensors. 
%{\partial^2 \over \partial t^2} - 
%\triangle=({\epsilon_{ijk} \partial_j p_k 
%\epsilon_{ilm} \partial\\epsilon_{ijk}_l p_m + 
%{\bf p}\cdot {\boldsymbol{\partial})/{\bf p}^2.
%$ 
We consider the same gauge functions 
$\chi_{nm}$.  
Now we can write the gauge covariant derivatives as 
\begin{equation}
D_\mu=\partial_\mu -ig_c A_\mu,
\label{GCD}
\end{equation} 
and thus the gauge transformation of a field $\phi(r)$ 
and $A$ are given as usual 
\begin{equation}
\phi(r)\rightarrow e^{ig_c\chi}\phi(r), \ \ \ 
A_\mu \rightarrow A_\mu +\partial_\mu \chi.
\label{GT}
\end{equation} 
%where $\chi$ stands for arbitrary linear combination of $\chi_{nm}$. 
Note here that in these gauge transformations $A_0$ does not change at all, 
because $\partial_t \chi=0$. 
In the present choice for $\chi$ given 
in Eq.~\eqref{chi} we have a trivial relation 
$$
[{1 \over c^2}{\partial^2 \over \partial t^2} - \vtri(x,y,z)]\chi 
=-\partial^\mu A_\mu.
$$ 
This fact means that by regauging $A_\mu$ in terms of $\chi_A=-zA_z$ 
we can get Lorentz 
condition $\partial^\mu A_\mu'=0$ where 
$A_\mu=A_\mu+\partial_\mu \chi_A$, but we find out $A_\mu'=0$ for 
$\forall \mu$ by choosing $A_0=0$. 
That is to say, we can eliminate the fields $A_\mu$ from the covariant 
derivatives, and thus all the effects can be described by the factors 
$e^{ig_c\chi_A}$ that are not the phase factor because $\chi_A$ are 
generally not real. 
$\chi_A$ has the infinite variaty arising from the degeneracy of the 
zero-energy states. 
How can we determine the gauge functions $\chi_A$? 
We can point out one possibility that the boudary conditions determine 
the gauge functions $\chi_A$. 
In this view the improper choice of the gauge functions $\chi_A$ for the 
boundary conditions produces the gauge interactions by $A_\mu$.  

Here we study the simplest case of $a=1$, where the equation is 
expressed by 
\begin{equation} 
[{1 \over c^2}{\partial^2 \over \partial t^2} - \vtri(x,y,z)-
g_1]\psi_z(r)=0.
 \label{tach}%\label{tach}
\end{equation} 
Taking account of $g_1> 0$, this equation is nothing but that for 
a tachyon having the negative squared-mass $m^2=-g_1$. 
This fact means that the tachyon can be a kind of these gauge fields and 
all the effects due to the tachyon are expressed 
by the factors $e^{ig_c\chi_A}$. 
% that are, of course, not the phase factors in general. 
Now we may say that in the world occupied by tachyons the vacuum will 
have the tachyon factors and then the space is 
possibly not flat any more. 

Here we shall briefly comment on massive particles 
apart from the gauge fields. 
In the case for $g_1=0$ 
Eq.~\eqref{tach} represents the equation for massless particles. 
The addition of $(-m^2+m^2)/\hbar^2$ to the potential term 
does not change the equation. 
We can, however, separate the equation into two parts 
for the fields $\phi({\bf r},t)=A_z(x,y)\psi(z,t)$ such that 
\begin{equation} 
[{1 \over c^2}{\partial^2 \over \partial t^2} -
{\partial^2 \over \partial z^2} +(m/\hbar)^2]\psi_z(z,t)=0
\label{FM}
\end{equation} 
and 
\begin{equation} 
[-\vtri(x,y) -(m/\hbar)^2]A_z(x,y)=0.
\label{FN}
\end{equation} 
One for the two dimensional $zt$ space is the usual equation 
for relativistic free motions for massive particles, 
whereas the other is nothing but 
the equation for the zero-energy states. 
Now we have found out a new freedom in the equation for massless particles 
in Lorenzt spaces, which express free motions of massive particles. 
Of course, such massive particles must have internal structures 
%described by $A_z(x,y)$ in the above case, 
which are described by the infinite degrees of freedom 
due to the zero-energy states. 
We should also note 
that the addition of a mass term $(m/\hbar)^2$ to Eq.~\eqref{rel} 
derive the equation for the massive particle which has 
internal structures of the zero-energy states for the 
internal potential $V_a$. 

Now we briefly study two other 
interesting gauges for $a=1/2$ and $a=3/2$. 
\hfil\break 
(1) For the $a=1/2$ case we have the potential 
$V_{1/2}(\rho)=-g_{1/2}\rho^{-1}$. 
This potential seems to represent a U(1)-type potential 
%(possibly gravitational one) 
between two particles %with opposite charges 
being in the $xy$ plane. 
Actually such situations will occur in the case 
for describing the currents in a surface.  
We may say that the fields represent the currents exchanged 
between two charged particles. % or possibly do gravitational fields. 
Note here that the current of the gauge fields %$A_z(x,y)$ 
in the $xy$ plane 
is described by the stationary flows turning over 
at the origin as shown in Figs.1 and 2, 
where $\alpha=0$ in Eq.~\eqref{psi} is taken, 
and the current is 
defined as usual 
$ j_\mu={\rm Re}[\phi^*(-i\hbar\partial_\mu \phi]$
~\cite{sk-jp,k-1,ks-pr,sk-nc}. 
These behaviors of the currents are easily understood, 
considering the fact that the gauge fields 
can be written by the solutions of Eq.~\eqref{psi} 
which are similar to the plane waves in the $u_{1/2}v_{1/2}$ plane. 
It is noted that the whole of the $u_{1/2}v_{1/2}$ plan
is represented by two sheets of the usaul $xy$ plane just as same as 
Riemann surface with the cut on the $x$ axis . 
We can understand the reason why the cuts exist in Figs.1 and 2, because  
the direction of the flows in the half-plane upper than the cuts 
is different from that of the lower half-plane.  
We may say that the sources of the gauge fields stay at origin in Figs.1 
and 2. 
The connection of the two flows given in Figs. 1 and 2 
can be performed as Fig.3. 
In this figure the gauge fields are bounded 
around the finite cut, 
and thus we may say that 
this state represents a bound state of two sources 
having opposite charges. 
Note here that we can easily make standing wave states without any cuts 
in the $u_av_a$ plane 
for arbitrary $a$~\cite{k-bio,k-lis,k-slow}, and those states can be 
mapped into the $xy$ plane 
in terms of the inverse transformations of the conformal transformations. 
\hfil\break 
(2) For the case of $a=3/2$ 
the potential has the linear dependence with respect 
to $\rho$. 
This situation reminds us of the SU(3) color potential introduced 
in quark-confinement dynamics. 
From the relations $\varphi_a=a \varphi$ 
we can generally show that 
the currents for the plane waves in the $u_av_a$ plane 
are described by the corner flows which round the center with the angle 
$\pi/a$ for arbitrary numbers of $a$~\cite{ks-pr}. 
For $a=3/2$ the current of the gauge fields in the $xy$ plane 
is described as Fig. 4 for $\alpha=0$, 
where three different currents must exist in the $xy$ plane 
because the angle of the corner flows is $2\pi/3$, 
and the cut again appears 
on the $x$ axis from 0 to $\infty$. 
It should be noticed that the cut appears only on one direction such as 
on the positive part of the $x$ axis in Fig.4. 
In this case we have two different types of the connections among the 
gauge fields given in Figs.5 and 6. 
It is obvious that Fig.6 represents a three gauge field vertex 
in non-Abelian gauge theories. 
We may consider that Fig.5 represents a diagram related to bound states of 
quark-antiquark like mesons in QCD, whereas Fig.6 does those for three 
quarks like baryons. 
To study the problems we have to understand the properties of sources 
that appear at the end of the cuts. 
Note finally that SU(N) gauge fields will appear for $a=N/2$, where $N=$ 
positive integers, and 
the cuts appear only in the cases for 
$N=$ odd integers.

\pagebreak

\pagebreak

\begin{figure}%[pb]
   \begin{center}
    \begin{picture}(200,200)
     \thicklines
    
     \put(0,100){\line(1,0){100}}
     \put(100,101){\line(1,0){100}}
     \put(100,99){\line(1,0){100}}
     \put(100,0){\vector(0,1){200}}
     \put(105,88){$0$}
     \put(205,98){$x$}
     \put(98,205){$y$}
     \put(98.6,97.6){$\bullet$}
     \put(130,105){${\rm cut}$}
 
     \put(102,120){\line(1,0){88}}
     \put(200,120){\vector(-1,0){10}}
     \put(102,80){\vector(1,0){98}}
     \qbezier(102,120)(80,100)(102,80)
     
     %\put(145,118){$\leftarrow$}
     %\put(145,78){$\rightarrow$}
     
     \put(200,100){\vector(1,0){3}}
     %\put(95.5,200){\vector(0,1){1}}
     %\put(95.5,0){\vector(0,-1){1}}
     %\put(105.5,0){\vector(0,-1){1}}
    \end{picture}
   \end{center}
   \caption[]{Flows of the states with the factor 
   $e^{-ik_{a}u_{a}}$ for $a=1/2$.}
   \label{fig:1a}
  \end{figure}

\begin{figure}%[pb]
 
    \begin{center}
    \begin{picture}(200,200)
     \thicklines
    
     \put(100,100){\vector(1,0){100}}
     \put(0,101){\line(1,0){100}}
     \put(0,99){\line(1,0){100}}
     \put(100,0){\vector(0,1){200}}
     \put(92,88){$0$}
     \put(205,98){$x$}
     \put(98,205){$y$}
     \put(98,97.6){$\bullet$}
     \put(50,105){${\rm cut}$}
 
     \put(10,120){\line(1,0){88}}
     \put(0,120){\vector(1,0){10}}
     \put(98,80){\vector(-1,0){98}}
     \qbezier(98,120)(120,100)(98,80)

    \end{picture}
   \end{center}
   \caption[]{Flows of the states with the factor 
   $e^{ik_{a}u_{a}}$ for $a=1/2$.}
   \label{fig:1b}
  \end{figure}

\begin{figure}%[pb]
   \begin{center}
    \begin{picture}(200,200)
     \thicklines
    
     %\put(0,100){\vector(1,0){200}}\put(50,105){${\rm cut}$}
     \put(50,101){\line(1,0){100}}
     \put(50,99){\line(1,0){100}}
     %\put(100,0){\vector(0,1){200}}
     %\put(105,88){$0$}
     %\put(205,98){$x$}
     %\put(98,205){$y$}
     \put(48.6,97.7){$\bullet$}
     \put(149,97.6){$\bullet$}
     \put(90,105){${\rm cut}$}
 
     \put(52,120){\line(1,0){48}}
     \put(148,120){\vector(-1,0){48}}
     \put(52,80){\vector(1,0){48}}
     \put(100,80){\line(1,0){48}}
     \qbezier(52,120)(30,100)(52,80)
     \qbezier(148,120)(170,100)(148,80)
     
     %\put(145,118){$\leftarrow$}
     %\put(145,78){$\rightarrow$}
     
     %\put(105.5,200){\vector(0,1){1}}
     %\put(95.5,200){\vector(0,1){1}}
     %\put(95.5,0){\vector(0,-1){1}}
     %\put(105.5,0){\vector(0,-1){1}}
    \end{picture}
   \end{center}
   \caption[]{Connection of two flows for $a=1/2$.}
   \label{fig:1c}
  \end{figure}

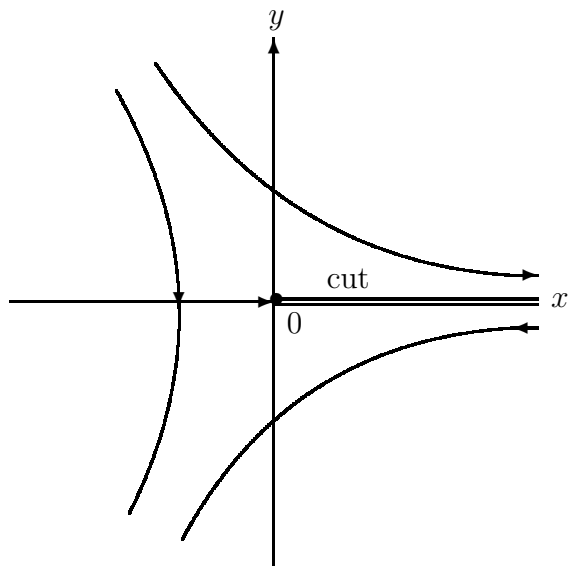
\begin{figure}%[pb]
   \begin{center}
    \begin{picture}(200,200)
     \thicklines
    
     \put(0,100){\vector(1,0){100}}
     \put(100,101){\line(1,0){100}}
     \put(100,99){\line(1,0){100}}
     \put(100,0){\vector(0,1){200}}
     \put(105,88){$0$}
     \put(205,98){$x$}
     \put(98,205){$y$}
     \put(98,98){$\bullet$}
     \put(120,105){${\rm cut}$}
 
     \put(190,110){\vector(1,0){10}}
     \qbezier(55,190)(105,115)(190,110)
     
     \put(64,102.5){\vector(0,-1){5}}
     \qbezier(40,180)(85,100)(45,20)
     
     \put(200,90){\vector(-1,0){10}}
     \qbezier(65,10)(105,85)(190,90)
     
    \end{picture}
   \end{center}
   \caption[]{Flows for $a=3/2$.}
   \label{fig:2a}
  \end{figure}

  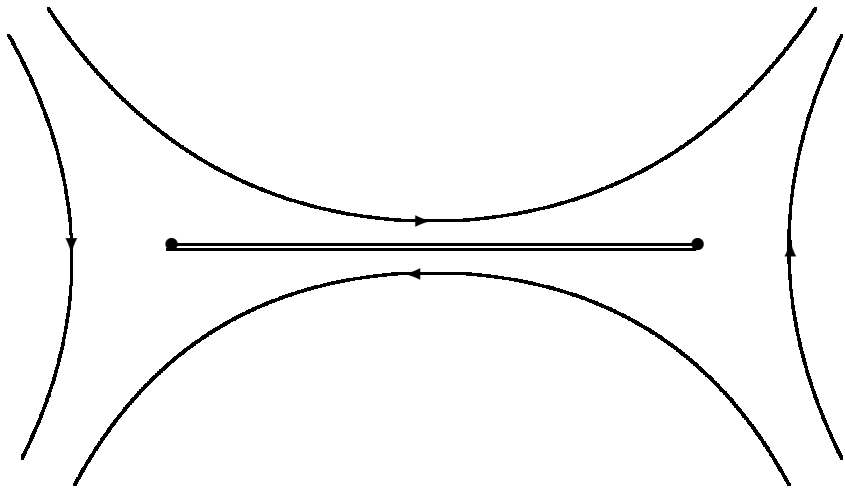
\begin{figure}%[pb]
   \begin{center}
    \begin{picture}(300,200)
     \thicklines
    
     %\put(0,100){\line(1,0){100}}
     \put(50,101){\line(1,0){200}}
     \put(50,99){\line(1,0){200}}
     \put(49,98){$\bullet$}
     \put(248,98){$\bullet$}
     %\put(205,98){$x$}
     %\put(98,205){$y$}
 
     \put(140,110){\vector(1,0){10}}
     \qbezier(5,190)(55,115)(140,110)
     \put(150,110){\line(1,0){10}}
     \qbezier(160,110)(245,115)(295,190)
     
     \put(14,102.5){\vector(0,-1){5}}
     \qbezier(-10,180)(35,100)(-5,20)
     \put(286,97.5){\vector(0,1){5}}
     \qbezier(305,180)(265,100)(305,20)
     
     \put(150,90){\vector(-1,0){10}}
     \qbezier(15,10)(55,85)(140,90)
     \put(150,90){\line(1,0){10}}
     \qbezier(160,90)(245,85)(285,10)

    \end{picture}
   \end{center}
   \caption[]{Connection of two flows for $a=3/2$.}
   \label{fig:2b}
  \end{figure}

  \begin{figure}%[pb]
   \begin{center}
    \begin{picture}(300,300)
     \thicklines
    
     %\put(0,100){\line(1,0){100}}
     \put(165,151){\line(1,0){100}}
     \put(165,149){\line(1,0){100}}
     \put(165,151){\line(-1,2){40}}
     \put(163,150){\line(-1,2){40}}
     \put(163,150){\line(-1,-2){40}}
     \put(165,149){\line(-1,-2){40}}
     \put(263,147.8){$\bullet$}
     \put(121.3,68){$\bullet$}
     \put(120.9,230){$\bullet$}
     %\put(100,0){\vector(0,1){200}}
     %\put(105,88){$0$}
     %\put(205,98){$x$}
     %\put(98,205){$y$}
   
       \put(260,170){\vector(-1,0){82}}
       \put(178,170){\vector(-1,2){35}}
       \put(143,240){\vector(1,2){30}}
       \put(290,230){\vector(-1,-2){30}}
       
       \put(178,130){\vector(1,0){80}}
       \put(143,60){\vector(1,2){35}}
       \put(258,130){\vector(1,-2){25}}
       \put(168,9.3){\vector(-1,2){25}}
       
       \put(97.5,220){\vector(1,-2){35}}
       \put(132,150){\vector(-1,-2){34}}
       \put(37.5,220){\vector(1,0){60}}
       \put(97.5,82){\vector(-1,0){60}}

     %\put(190,110){\vector(1,0){10}}
     \qbezier(317,220)(272,150)(311,95)
     \put(293.5,152.5){\vector(0,-1){5}}
     \qbezier(144,-2)(109,55)(33,55)
     \put(108,36){\vector(-1,1){3}}
     \qbezier(140,306)(112,248)(34,245)
     \put(105,265){\vector(1,1){2.5}}
     %\qbezier(0,90)(90,72.8)(140,-23.2)

     %\put(64,102.5){\vector(0,-1){5}}
     %\qbezier(40,180)(85,100)(45,20)
     
     %\put(200,90){\vector(-1,0){10}}
     %\qbezier(65,10)(105,85)(190,90)

    \end{picture}
   \end{center}
   \caption[]{Image for 
   the connection of three flows for $a=3/2$, 
   where all the angles between two connecting vectors 
   should be $3\pi/2$ 
   in real connections.}
   \label{fig:2.c}
  \end{figure}
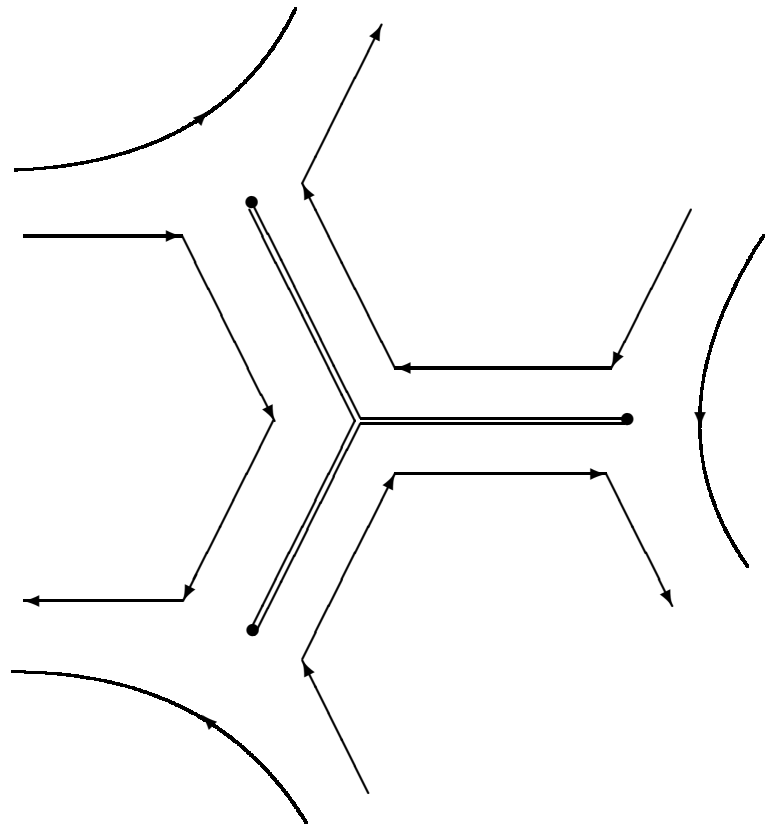

\end {document}